\documentclass[12pt]{elsarticle}
\usepackage{amsmath,amssymb}
\usepackage{eepic}
\usepackage{hyperref}
\usepackage{graphicx}
\usepackage{enumitem}\biboptions{sort&compress}
\usepackage[left=2.0cm,right=2.0cm,top=1cm,bottom=2cm,bindingoffset=0cm]{geometry}
\usepackage{color}
\renewcommand{\alarm}[1]{\textcolor{cyan}{#1}} %cyan

\newtheorem{lemma}{Lemma}
\newtheorem{theorem}{Theorem}
\newtheorem{proposition}{Proposition}
\newtheorem{corollary}{Corollary}
\newtheorem{definition}{Definition}

\journal{Physics Letters A [accepted] doi: \href{http://dx.doi.org/10.1016/j.physleta.2016.04.036}{10.1016/j.physleta.2016.04.036}}

\DeclareMathOperator{\LCE}{LCE}
\DeclareMathOperator{\LEs}{LE}

\begin{document}

\begin{frontmatter}

\title{
  The Lyapunov dimension and its estimation via the Leonov method.
}

 \author{N.V. Kuznetsov\corref{cor}\,}
 \ead{nkuznetsov239@gmail.com}

 \address[spbu]{Faculty of Mathematics and Mechanics,
 St.~Petersburg State University, Russia}
 \address[fin]{Department of Mathematical Information Technology,
 University of Jyv\"{a}skyl\"{a}, Finland}

\begin{abstract}
  Along with widely used numerical methods for estimating and computing
  the Lyapunov dimension there is an effective analytical approach,
  proposed by G.A.~Leonov in 1991.
  The Leonov method is based on the direct Lyapunov method
  with special Lyapunov-like functions.
  The advantage of the method
  is that it allows one to estimate the Lyapunov dimension
  of invariant sets without localization of the set in the phase space
  and, in many cases, to get effectively an exact Lyapunov dimension formula.

  In this work the invariance of the Lyapunov dimension
  with respect to diffeomorphisms and its connection with the Leonov method are discussed.
  For discrete-time dynamical systems an analog of Leonov method is suggested.
  In a simple but rigorous way,
  here it is presented the connection between the Leonov method
  and the key related works:
  Kaplan and Yorke (the concept of the Lyapunov dimension, 1979),
  Douady and Oesterl\'{e}
  (upper bounds of the Hausdorff dimension via the Lyapunov dimension of maps, 1980),
  Constantin, Eden, Foia\c{s}, and Temam
  (upper bounds of the Hausdorff dimension via
  the Lyapunov exponents and Lyapunov dimension of dynamical systems, 1985-90),
  and the numerical calculation of the Lyapunov exponents and dimension.
 \end{abstract}


\end{frontmatter}

\section{Introduction}
The concept of the Lyapunov dimension was suggested in the seminal paper
by Kaplan and Yorke \cite{KaplanY-1979} for estimating the Hausdorff dimension
of attractors. The direct numerical computation
of the Hausdorff dimension
of attractors is often a problem of high numerical complexity
(see, e.g. discussion in \cite{RusselHO-1980}),
thus, various estimates of this dimension are of interest.
Later the concept of the Lyapunov dimension has been developed
in a number of papers
(see, e.g. \cite{Ledrappier-1981,FredericksonKYY-1983,KaplanMY-1984,GrassbergerP-1983,ConstantinFT-1985,Smith-1986,EdenFT-1991,Hunt-1996} and others).

Along with widely used numerical methods for estimating and computing
the Lyapunov dimension
there is an effective analytical approach, proposed by Leonov in 1991 \cite{Leonov-1991-Vest}
(see also \cite{LeonovB-1992,Leonov-2002,Leonov-2012-PMM,LeonovK-2015-AMC,LeonovKM-2015-EPJST}).
The Leonov method is based on the direct Lyapunov method
with special Lyapunov-like functions.
The advantage of Leonov method is that it allows one to estimate the Lyapunov dimension
of invariant sets without localization of the set in the phase space
and in many cases to get
effectively
exact Lyapunov dimension formula
\cite{Leonov-2002,LeonovP-2005,LeonovPS-2011,LeonovK-2015-AMC,LeonovKKK-2015-arXiv-YangTigan,LeonovAK-2015,LeonovKKK-2016-CNSCS,LeonovKM-2015-ArXiv}.

Further the invariance of the Lyapunov dimension
with respect to diffeomorphisms and its connection with the Leonov method are discussed.
For discrete-time dynamical systems an analog of Leonov method is suggested.
%Some supplementary important facts,
%which are not directly used in the presentation,
%are given as footnotes.

\section{Lyapunov dimension of maps and dynamical systems}
Consider an autonomous differential equation
\begin{equation} \label{eq:ode}
 \dot u = f(u), \quad f : U \subseteq \mathbb{R}^n \to \mathbb{R}^n,
\end{equation}
where $f$ is a continuously differentiable vector-function.
Suppose that any solution $u(t,u_0)$ of \eqref{eq:ode} such that $u(0,u_0)=u_0 \in U$
exists for $t \in [0,\infty)$, it is unique and stays in $U$.
Then the evolutionary operator $\varphi^t(u_0) = u(t,u_0)$
is continuously differentiable and satisfies the semigroup property:
\begin{equation}\label{group_prop}
    \varphi^{t+s}(u_0) = \varphi^{t}(\varphi^{s}(u_0)), \ \varphi^0(u_0)=u_0
    \quad \forall ~ t, s \geq 0, \ \forall u_0 \in U.
\end{equation}
Thus, $\{\varphi^t\}_{t\geq0}$ is a smooth \emph{dynamical system}
in the phase space $(U, ||~\cdot~||)$:
$\big(\{\varphi^t\}_{t\geq0},(U \subseteq \mathbb{R}^n,||\cdot||) \big)$.
Here $||u|| = \sqrt{u_1^2 + \cdots + u_n^2}$
is Euclidean norm of the vector
${u} = (u_1, \ldots, u_n) \in \mathbb{R}^n$.
Similarly, one can consider a dynamical system generated by the difference equation
\begin{equation}\label{eq:odife}
   u(t+1) = \varphi(u(t)), \quad t = 0,1,..\,,
\end{equation}
where $\varphi : U \subseteq \mathbb{R}^n \to U$ is a continuously differentiable vector-function.
Here
$\varphi^t(u) = \underbrace{(\varphi\circ \varphi \circ \cdots \varphi)(u)}\limits_{t\ {\rm times}}$, $\varphi^0(u)=u$,
and the existence and uniqueness (in the forward-time direction)
take place for all $t\geq0$.
Further $\{\varphi^t\}_{t\geq0}$ denotes a smooth \emph{dynamical system} with continuous or discrete time.

Consider the linearizations of systems (\ref{eq:ode}) and \eqref{eq:odife}
along the solution $\varphi^t(u)$:
\begin{equation} \label{sflct}
  \dot y = J(\varphi^t(u))y,  \quad J(u) = Df(u),\\
\end{equation}
\begin{equation} \label{sfldt}
  y(t+1) = J(\varphi^t(u))y(t), \quad J(u) = D\varphi(u),
\end{equation}
where $J(u)$ is the $n\times n$ Jacobian matrix,
the elements of which are continuous functions of $u$.
Suppose that $\det J(u)\neq0$ $\forall u \in U$.

Consider the fundamental matrix, which consists of linearly independent solutions $\{y^i(t)\}_{i=1}^{n}$ of the linearized system,
\begin{equation}\label{nfm}
  D\varphi^t(u)=\big(y^1(t),...,y^n(t)\big),
  \quad D\varphi^0(u) = I,
\end{equation}
where $I$ is the unit $n\times n$ matrix.
An important cocycle property of fundamental matrix \eqref{nfm} is as follows
\begin{equation}\label{cocycle}
  D\varphi^{t+s}(u) = D\varphi^t\big(\varphi^s(u)\big)D\varphi^{s}(u),
  \ \forall t,s \geq 0, \ \forall u \in U.
\end{equation}

Let $\sigma_i(t,u)=\sigma_i(D\varphi^t(u))$, $i=1,2,..,n$, be the singular values of $D\varphi^t(u)$
(i.e. $\sigma_i(t,u)>0$ and $\sigma_i(t,u)^2$ are the eigenvalues of the symmetric matrix $D\varphi^t(u)^*D\varphi^t(u)$ with respect to their algebraic multiplicity),
ordered so that
$\sigma_1(t,u)\geq \cdots \geq \sigma_n(t,u) > 0$ for any $u \in U$, $t\geq0$.
The singular value function of order $d \in [0,n]$ at the point $u \in U$
for $D\varphi^t(u)$ is defined as
\begin{equation}\label{defomega}
  \omega_d(D\varphi^t(u)) = \left\{
  \begin{aligned}
    & 1, && d=0, \\
    & \sigma_1(t,u)\sigma_2(t,u)\cdots\sigma_{d}(t,u), && d \in \{1,2,..,n\}, \\ %to  avoid 0^0
    & \sigma_1(t,u)\cdots\sigma_{\lfloor d \rfloor}(t,u)\sigma_{\lfloor d \rfloor+1}(u)^{d-\lfloor d \rfloor}, && d \in (0,n),
  \end{aligned}
  \right.
\end{equation}
where ${\lfloor d \rfloor}$ is the largest integer less or equal to $d$.
Remark that $|\det D\varphi^t(u)|=\omega_n(D\varphi^t(u))$.
Similarly, we can introduce the singular value function for arbitrary quadratic matrices
and then by the Horn inequality \cite{HornJ-1994-book}
for any two $n\times n$ matrices $A$ and $B$ and any $d \in [0,n]$
we have (see, e.g. \cite[p.28]{BoichenkoLR-2005})
\begin{equation}\label{horn}
  \omega_d(AB) \leq \omega_d(A)\omega_d(B), \quad d \in [0,n].
\end{equation}

Let a nonempty set $K \subset U \subseteq \mathbb{R}^n$
be invariant with respect to the dynamical system $\{\varphi^t\}_{t\geq0}$,
i.e. $\varphi^t(K)=K$ for all $t>0$.
Since in the numerical experiments only finite time $t$ can be considered,
for a fixed $t \geq 0$ let us consider the map defined by the evolutionary operator $\varphi^t(u)$: $\varphi^t : U \subseteq \mathbb{R}^n \to U$.

The concept of the Lyapunov dimension was suggested in the seminal paper
by Kaplan and Yorke \cite{KaplanY-1979} and
later it was developed in a number of papers
(see, e.g. \cite{Ledrappier-1981,FredericksonKYY-1983,ConstantinFT-1985,Hunt-1996,Eden-1990}).
The following definition is inspirited by Douady and Oesterl\'{e} \cite{DouadyO-1980}.
\begin{definition}\label{def:DOlocal-map}
The local Lyapunov dimension\footnote{This is not a dimension
in a rigorous sense (see, e.g. \cite{HurewiczW-1941,Kuratowski-1966,AleksandrovP-1973}).
The notion '\emph{local Lyapunov dimension}' is used, e.g. in \cite{Eden-1989,Hunt-1996}.}
of the map $\varphi^t$
(or finite-time local Lyapunov dimension of the dynamical system $\{\varphi^t\}_{t\geq0}$)
at the point $u \in U$ is defined as %(see, e.g.)
\begin{equation}\label{defLDmaplocal}
  \dim_{\rm L}(\varphi^t,u) =
  \inf\{d \in [0,n]: \omega_{d}(D\varphi^t(u)) < 1\}.
\end{equation}
If the infimum is taken over an empty set (i.e. $\omega_{n}(D\varphi^t(u)) \geq 1$),
we assume that the infimum and considered dimension
are taken equal\,\,\footnote{
In general, since $\omega_{0}(D\varphi^t(u))\equiv 1$ and
$d \mapsto \omega_{d}(D\varphi^t(u))$ is a left-continuous function,
we have
$\dim_{\rm L}(\varphi^t,u)
=
\max\{d \in [0,n]: \omega_{d}(D\varphi^t(u)) \geq 1\}$.
If all $\{\sigma_i(t,u)\}_1^n$ are assumed to be positive and $\omega_{n}(D\varphi^t(u)) < 1$,
then in \eqref{defLDmaplocal} the infimum is achieved
(see \eqref{lnomegat=0} and the Kaplan-Yorke formula \eqref{lftKY}).}
 to $n$.

The Lyapunov dimension of the map $\varphi^t$
(or finite-time Lyapunov dimension of the dynamical system $\{\varphi^t\}_{t\geq0}$)
with respect to the invariant set $K$
is defined as
\begin{equation}\label{DOmaptmax}
    \dim_{\rm L}(\varphi^t,K) = \sup\limits_{u \in K} \dim_{\rm L}(\varphi^t,u)
    = \sup\limits_{u \in K} \inf\{d \in [0,n]: \omega_{d}(D\varphi^t(u)) < 1 \}.
\end{equation}
\end{definition}

The continuity of the functions
$u \mapsto \sigma_i(D\varphi^t(u))$, $i=1,2,..,n$, on $U$
implies that
for any $d \in [0,n]$ and $t\geq0$
the function $u \mapsto \omega_{d}(D\varphi^t(u))$
is continuous on $U$ (see, e.g. \cite{DouadyO-1980},\cite[p.554]{Gelfert-2003}).
Therefore for a compact set $K \subset U$ and $t\geq0$ we have
\begin{equation}\label{wsupmax}
    \sup_{u\in K}\omega_d(D\varphi^{t}(u))=\max_{u\in K}\omega_d(D\varphi^{t}(u)).
\end{equation}
By relation \eqref{wsupmax} for a compact invariant set $K$ one can prove that
\begin{equation}\label{DOmapt}
   \dim_{\rm L}(\varphi^t,K)=\inf\{d\in[0,n]: \max\limits_{u \in K}\omega_{d}(D\varphi^t(u))<1\}.
\end{equation}

In the seminal paper \cite{DouadyO-1980} Douady and Oesterl\'{e}
proved rigorously that
the Lyapunov dimension of the map $\varphi^t$
with respect to the compact invariant set $K$
is an upper estimate of the Hausdorff dimension of the set $K$:
\begin{equation}\label{DO}
  \dim_{\rm H} K \leq \dim_{\rm L}(\varphi^t,K).
\end{equation}
For numerical estimations of dimension, the following remark is important.
From \eqref{cocycle} and \eqref{horn} it follows that
\[
  \sup_{u\in K}\omega_d(D\varphi^{t+s}(u)) =
  \sup_{u\in K}\omega_d\big(D\varphi^{t}(\varphi^{s}(u))D\varphi^{s}(u)\big)
  \leq
  \sup_{u\in K}\omega_d(D\varphi^{t}(u))
  \sup_{u\in K}\omega_d(D\varphi^{s}(u)) \quad \forall t,s\geq0
\]
and
$
\sup\limits_{u \in K}\omega_{d}(D\varphi^{nt}(u))\leq
(\sup\limits_{u \in K}\omega_{d}(D\varphi^{t}(u)))^n
$
for any integer $n\geq0$.
Thus for any $t\geq0$ there exists $\tau=\tau(t)>0$ such that
\begin{equation}\label{DOinctT}
  \dim_{\rm L}(\varphi^{t+\tau},K) \leq \dim_{\rm L}(\varphi^t,K).
\end{equation}

While in the computations we can consider only finite time $t$
and the map $\varphi^t$,
from a theoretical point of view,
it is interesting to study
the limit behavior %of the dynamical system $\{\varphi^t\}_{t\geq0}$
of finite-time Lyapunov dimension of the dynamical system $\{\varphi^t\}_{t\geq0}$
with respect to the compact invariant set $K$.
\begin{definition}\label{def:DOlocal-ds}
The Lyapunov dimension of the dynamical system $\{\varphi^t\}_{t\geq0}$
with respect to the invariant set $K$ is defined as
\begin{equation}
  \dim_{\rm L}(\{\varphi^t\}_{t\geq0},K) = \inf_{t > 0}\dim_{\rm L}(\varphi^t,K).
\end{equation}
\end{definition}
From \eqref{DO} and \eqref{DOmapt} we have
\begin{equation}\label{DOinf}
  \dim_{\rm H} K \leq \dim_{\rm L}(\{\varphi^t\}_{t\geq0},K) =
  \inf_{t > 0} \sup_{u \in K} \dim_{\rm L}(\varphi^t,u)
\end{equation}
and \eqref{DOinctT} implies
\begin{equation}\label{DOlim}
  \inf_{t > 0}\dim_{\rm L}(\varphi^t,K) = \liminf_{t \to +\infty}\dim_{\rm L}(\varphi^t,K).
\end{equation}

Remark that if $\sup_{u \in K}\omega_{\overline{d}}(D\varphi^t(u))<1$ 
for a certain $\overline{d} \in [0,n]$,
then
\begin{equation}\label{liminf}
  \inf_{t > 0}\sup\limits_{u \in K}\omega_{\overline{d}}(D\varphi^t(u))=
  \liminf_{t \to +\infty}\sup\limits_{u \in K}\omega_{\overline{d}}(D\varphi^t(u))=0
\end{equation}
and
\begin{equation}\label{dimLest}
  \dim_{\rm L}(\{\varphi^t\}_{t\geq0},K) \leq \dim_{\rm L}(\varphi^t,K) < \overline{d}.
\end{equation}

\begin{definition}\label{defLE} (see, e.g. \cite{AbarbanelBK-1991})
The finite-time Lyapunov exponents
(or the Lyapunov exponent functions of singular values)
of the dynamical system $\{\varphi^t\}_{t\geq0}$
at the point $u \in U$ are denoted by
\(
   \LEs_i(t,u) = \LEs_i(D\varphi^t(u)),\ i=1,2,..,n,
\)
and defined as
\[
  \LEs_i(t,u) = \frac{1}{t}\ln\sigma_i(t,u), \quad t>0.
\]
\end{definition}
Here $\LEs_1(t,u) \geq\cdots\geq\LEs_n(t,u)$ for all $t>0$ since
the singular values are ordered by decreasing.

For the sake of simplicity,
we assume that $\omega_1(D\varphi^t(u))>1>\omega_n(D\varphi^t(u))$
for $t>0, u \in K$.
Thus, $n>\dim_{\rm L}(\varphi^t,u)>1$ and
$\omega_{\dim_{\rm L}(\varphi^t,u)}(D\varphi^t(u))=1$.
Therefore for $j(t,u) = \lfloor \dim_{\rm L}(\varphi^t,u) \rfloor$ and
$s(t,u) = \dim_{\rm L}(\varphi^t,u)-\lfloor \dim_{\rm L}(\varphi^t,u) \rfloor$
we have
\begin{equation}\label{lnomegat=0} % avoid 0^0
  0=\frac{1}{t}\ln(\omega_{j(t,u)+s(t,u)}(D\varphi^t(u))) =
  \sum_{i=1}^{j(t,u)}\LEs_i(t,u) + s(t,u)\LEs_{j(t,u)+1}(t,u).
\end{equation}
Since $\LEs_i(t,u)$ are ordered by decreasing and $s(t,u)<1$, we have
\begin{equation}
\begin{aligned}\label{defjsut}
  &
  j(t,u) = \max\{m: \sum_{i=1}^{m}\LEs_i(t,u) \geq 0\},
  \quad \sum_{i=1}^{j(t,u)+1}\LEs_i(t,u)<0,\quad \LEs_{j(t,u)+1}(t,u)<0, \\
  &
  0 \leq s(t,u) =
     \dfrac{\LEs_1(t,u)+\cdots+\LEs_{j(t,u)}(t,u)}{|\LEs_{j(t,u)+1}(t,u)|} < 1.
\end{aligned}
\end{equation}
If $j(t,u)=0$ or $j(t,u)=n$, then let $s(t,u)=0$.
The expression
\begin{equation}\label{lftKY}
  \dim_{\rm L}^{\rm KY}(\{\LEs_i(t,u)\}_{i=1}^{n}) =j(t,u)+\dfrac{\LEs_1(t,u)+\cdots+\LEs_{j(t,u)}(t,u)}{|\LEs_{j(t,u)+1}(t,u)|}
\end{equation}
corresponds to the Kaplan-Yorke formula \cite{KaplanY-1979}
with respect to finite-time Lyapunov exponents
(the set $\{\LEs_i(t,u)\}_{1}^{n}$, ordered by decreasing\footnote{
   Various characteristics of chaotic behavior
   are based on the limit values of finite-time Lyapunov exponents (LEs):
   $\LEs_i(u) = \limsup\limits_{t \to +\infty} \LEs_i(t,u), i=1,..,n$.
   For example, Kaplan-Yorke formula with respect to LEs is considered
   in \cite{ConstantinFT-1985,EdenFT-1991}
   and the sum of positive LEs may be used \cite{Millionschikov-1976,Pesin-1977}
   as the characteristic of Kolmogorov-Sinai entropy rate (see \cite{Kolmogorov-1959,Sinai-1959,AdlerKA-1965,Dinaburg-1971}).
   Relying on ergodicity,
   the LEs and Lyapunov dimension of attractor are often computed along one trajectory
   (see also \cite{KaplanY-1979,Ledrappier-1981,FredericksonKYY-1983,FarmerOY-1983}),
   which is attracted or belongs to the attractor.
   But, in general,
   one has to consider a grid of points on $K$ and
   the corresponding local Lyapunov dimensions (see, e.g. \cite{KuznetsovMV-2014-CNSNS,LeonovKM-2015-EPJST}).
   For a given invariant set $K$ and a given point $u_0 \in K$
   there are two essential questions related
  to the computation of the Lyapunov exponents and
  the use of the Kaplan-Yorke formulas of local Lyapunov dimension
  $\sup_{u \in K}\dim_{\rm L}^{\rm KY}(\{\limsup_{t\to+\infty}\LEs_i(t,u)\}_1^n)$:
  whether $\limsup\limits_{t\to+\infty}\LEs_i(t,u_0) = \lim\limits_{t\to+\infty}\LEs_i(t,u_0)$ is valid, and if not, whether the relation
  $
   \sup_{u \in K} \dim_{\rm L}^{\rm KY}(\{\LEs_m(u)\}_1^n)
   =
   \sup_{u \in K \backslash \{\varphi^t(u_0), t \geq 0 \}} \dim_{\rm L}^{\rm KY}(\{\LEs_i(u)\}_1^n)
   $
   is true.
  In order to get rigorously the positive answer to these questions,
  from a theoretical point of view, one may use
  various ergodic properties of the dynamical system $\{\varphi^t\}_{t\geq0}$
  (see, e.g. Oseledets~\cite{Oseledec-1968}, Ledrappier~\cite{Ledrappier-1981},
  and auxiliary results in \cite{BogoliubovK-1937,DellnitzJ-2002}).
  However, from a practical point of view,
  the rigorous use of the above results is a challenging task
  (e.g. even for the well-studied Lorenz system)
  and hardly can be done effectively in the general case
  (see, e.g. the corresponding discussions
  in \cite{BarreiraS-2000},\cite[p.118]{ChaosBook},\cite{OttY-2008},\cite[p.9]{Young-2013}
   and the works \cite{LeonovK-2007,KuznetsovL-2005}
  on the Perron effects of the largest Lyapunov exponent sign reversals).
  For an example of the effective rigorous use of the ergodic theory
  for the estimation of the Hausdorff and Lyapunov dimensions, see, e.g. \cite{Schmeling-1998}.
}).
Remark that there exists $\overline{s}$
such that $s(t,u)<\overline{s}<1$ and
$\omega_{j(t,u)+\overline{s}}(D\varphi^t(u))<1$ for any such $\overline{s}$,
and $\omega_{\underline{d}}(D\varphi^t(u)) \geq 1$ for $0\leq \underline{d}\leq j(t,u)+s(t,u)$.
Thus $j(t,u)+s(t,u)=\dim_{\rm L}(\varphi^t,u)$ for $j(t,u)$ and $s(t,u)$ defined by \eqref{defjsut}.
Therefore, we get
\begin{proposition}\label{thm:ftKY}
For the Lyapunov dimension of the map $\varphi^t$
(or finite-time Lyapunov dimension of the dynamical system $\{\varphi^t\}_{t\geq0}$)
with respect to the compact invariant set $K$ we have
%\begin{equation}\label{ftKY}
\[
 \dim_{\rm H}K \leq
 \sup_{u\in K}\dim_{\rm L}(\varphi^t,u) =
 \sup_{u\in K}\dim_{\rm L}^{\rm KY}(\{\LEs_i(t,u)\}_{1}^{n})=
 \sup_{u\in K}
 \left(
   j(t,u)+\dfrac{\LEs_1(t,u)+\cdots+\LEs_{j(t,u)}(t,u)}{|\LEs_{j(t,u)+1}(t,u)|}
 \right).
\]
%\end{equation}
\end{proposition}

For numerical computation of the finite-time Lyapunov exponents\footnote{
Another widely used definition of the Lyapunov exponents goes back to Lyapunov \cite{Lyapunov-1892}.
Finite-time Lyapunov exponents $\{\LCE_i(t,u)\}_1^n$  of the fundamental matrix columns
$(y^1(t,u),...,y^n(t,u))=D\varphi^t(u)$
(called also finite-time Lyapunov characteristic exponents, LCE)
are defined as the set
$\{\frac{1}{t}\ln||y^i(t,u)||\}_1^n$ ordered by decreasing for $t>0$.
In contrast to the definition of the Lyapunov exponents of singular values,
finite-time Lyapunov exponents of fundamental matrix columns
may be different for different fundamental matrices (see, e.g. \cite{KuznetsovAL-2016}).
To get the set of all possible values of the Lyapunov exponents
of fundamental matrix columns
(the set with the minimal sum of values),
one has to consider the so-called normal fundamental matrices \cite{Lyapunov-1892}.
Using, e.g, Courant-Fischer theorem \cite{HornJ-1994-book},
it is possible to show that
$\LCE_1(t,u) =\LEs_1(t,u)$
and $\LEs_i(t,u) \leq \LCE_i(t,u)$ for $1<i\leq n$, and, thus,
$\dim_{\rm L}^{\rm KY}(\{\LEs_i(t,u)\}_1^n) \leq \dim_{\rm L}^{\rm KY}(\{\LCE_i(t,u)\}_1^n)$.
For example, for the matrix \cite{KuznetsovAL-2016}
 \(
    X(t)=\left(
      \begin{array}{cc}
        1 & g(t)-g^{-1}(t) \\
        0 & 1 \\
      \end{array}
    \right)
 \)
 we have the following ordered values:
 $  \LCE_1(X(t)) =
  {\rm max}\big(\limsup\limits_{t \to +\infty}\frac{1}{t}\ln|g(t)|,
  \limsup\limits_{t \to +\infty}\frac{1}{t}\ln|g^{-1}(t)|\big),
  \LCE_2(X(t)) = 0$;
 $
  \LEs_{1,2}(X(t)) = {\rm max, min}
  \big(
     \limsup\limits_{t \to +\infty}\frac{1}{t}\ln|g(t)|,
     \limsup\limits_{t \to +\infty}\frac{1}{t}\ln|g^{-1}(t)|
  \big).
 $
 %Remark that for both diagonal elements
 %$\limsup\limits_{t \to +\infty}\frac{1}{t}\ln|1|=0$.

 \noindent The various generalizations of the Lyapunov exponents and their properties
 are studied, e.g., in %BylovVGN-1966,
 \cite{BylovVGN-1966,Pesin-1977,KunzeK-2001,LeonovK-2007,BarreiraG-2011,Izobov-2012,CzornikNN-2013}.
}
there are developed various continuous and discrete algorithms
based on QR and SVD decompositions of fundamental matrix
(see, e.g. MATLAB realizations in \cite{KuznetsovMV-2014-CNSNS,LeonovKM-2015-EPJST}).
However such algorithms may not work well in the case of coincidence or closeness
of two or more Lyapunov exponents. %and in the case of irregular linearizations
Also it is important to remark that numerical computation of the Lyapunov exponents
can be done only for a finite time $T$,
the justification of the choice of which is usually omitted,
while it is known that in such computations
unexpected ``jumps'' can occur (see, e.g. \cite[p.116, Fig.6.3]{ChaosBook}).
The various methods (see, e.g. \cite{WolfSSV-1985,RosensteinCL-1993,AbarbanelBST-1993,HeggerKS-1999})
are also developed for the estimation of the Lyapunov exponents from time series.
However there are known examples in which the results of such computations
differ substantially from the analytical results \cite{TempkinY-2007,AugustovaBC-2015}.

\section{
Invariance with respect to diffeomorphisms and analytical estimates}

While the topological dimensions are invariant with respect to Lipschitz homeomorphisms,
the Hausdorff dimension is invariant with respect to Lipschitz diffeomorphisms
and the noninteger Hausdorff dimension is not invariant with respect to
homeomor\-phisms \cite{HurewiczW-1941}.
Since the Lyapunov dimension is used as an upper estimate of the Hausdorff dimension,
the question arises whether the Lyapunov dimension
is invariant under diffeomorphisms
(see, e.g. \cite{OttWY-1984}).

Consider the dynamical system
$\big(\{\varphi^t\}_{t\geq0},(U\subseteq \mathbb{R}^n,||\cdot||) \big)$
under the change of coordinates $w = h(u)$,
where $h: U \subseteq \mathbb{R}^n \to \mathbb{R}^n$ is a diffeomorphism.
In this case
the semi-orbit
$\gamma^{+}(u) = \{\varphi^t(u), t \geq 0 \}$
is mapped to the semi-orbit defined by
$\varphi_h^t(w)=\varphi_h^t(h(u))=h(\varphi^t(u))$,
the dynamical system
$\big(\{\varphi^t\}_{t\geq0},(U\subseteq \mathbb{R}^n,||\cdot||) \big)$
is transformed to
the dynamical system
$\big(\{\varphi_h^t\}_{t\geq0},(h(U)\subseteq \mathbb{R}^n,||\cdot||) \big)$,
and the compact set $K \subset U$ invariant with respect to $\{\varphi^t\}_{t\geq0}$
is mapped to the compact set $h(K) \subset h(U)$
invariant with respect to $\{\varphi_h^t\}_{t\geq0}$.
Here
\[
  D_w\varphi_h^t(w)=
  D_w\big(h(\varphi^t(h^{-1}(w)))\big)=
  D_uh(\varphi^t(h^{-1}(w)))
  D_u\varphi^t(h^{-1}(w))
  D_w h^{-1}(w),
\]
\[
  D_u \big( \varphi_h^t(h(u)) \big)=
  D_w\varphi_h^t(h(u)) D_uh(u)=
  D_u\big(h(\varphi^t(u))\big)=
  D_u h(\varphi^t(u))D_u\varphi^t(u).
\]
Therefore
\[
  D_w h^{-1}(w) = \big(D_uh(u)\big)^{-1}
\]
and\begin{equation}\label{Dphih}
  D\varphi_h^t(w)=
  Dh(\varphi^t(u))
  D\varphi^t(u)
  \big(Dh(u)\big)^{-1}.
\end{equation}

If $u \in K$, then $\varphi^t(u)$ and $\varphi_h^t(h(u))$
define bounded semi-orbits.
Remark that $Dh$ and $(Dh)^{-1}$ are continuous and, thus,
$Dh(\varphi^t(u))$ and $(Dh(\varphi^t(u)))^{-1}$ are bounded in $t$.
From \eqref{wsupmax} it follows that
for any $d \in [0,n]$ there is a constant $c=c(d)\geq 1$
such that (see also \cite[p.29]{BoichenkoLR-2005})
\begin{equation}\label{Dhest}
\begin{aligned}
  &
  \max_{u \in K}\omega_d\big(Dh(u)\big)\leq c,
  \quad
  \max_{u \in K}\omega_d\big((Dh(u))^{-1}\big)
  \leq c.
\end{aligned}
\end{equation}

\begin{lemma}\label{thm:hdiff}
If for $t>0$  there exist diffeomorphism $h: U \subseteq \mathbb{R}^n \to \mathbb{R}^n$
and $d \in [0,n]$ such that the estimation\,\footnote{The expression in \eqref{wDphiht<1}
corresponds to the expressions considered in
\cite[eq.(1)]{Leonov-1991-Vest} for $p(u)=Dh(u)$,
\cite[eq.(1)]{Leonov-2002} and \cite[p.99, eq.10.1]{Leonov-2008} for $Q(u)=Dh(u)$.}
\begin{equation}\label{wDphiht<1}
  \max_{w \in h(K)}\omega_d\big(D\varphi_h^t(w)\big)
  =
  \max_{u \in K}
  \omega_d\bigg(
  Dh(\varphi^t(u))
  D\varphi^t(u)
  \big(Dh(u)\big)^{-1}
  \bigg) <1
\end{equation}
is valid,
then for $u \in K$ we get
 \[
    \liminf\limits_{t \to +\infty}
      \bigg(
      %\sup_{u \in K}
      \omega_d\big(D\varphi^t(u)\big)
      -
      %\sup_{w \in h(K)}
      \omega_d\big(D\varphi_h^t(h(u))\big)
      \bigg)=0
 \]
 and
 \[
  \liminf\limits_{t \to +\infty}
  %\sup_{u \in K}
  \omega_d\big( D\varphi_h^t(h(u)) \big)
  =
  \liminf\limits_{t \to +\infty}
  %\sup_{u \in K}
  \omega_d\big(D\varphi^t(u)\big)=0.
 \]
\end{lemma}
{\bf Proof.}
Applying \eqref{horn} to \eqref{Dphih}, we get
\[
  \omega_d\big(D\varphi_h^t(h(u))\big) \leq
  \omega_d\big(Dh(\varphi^t(u))\big)
  \omega_d\big(D\varphi^t(u)\big)
  \omega_d\big(\big(Dh(u)\big)^{-1}\big).
\]
By \eqref{Dhest} we obtain
\[
  %\sup_{u \in K}
  \omega_d\big(D\varphi_h^t(h(u))\big) \leq
  c^2
  %\sup_{u \in K}
  \omega_d\big(D\varphi^t(u)\big).
\]
Similarly,
\[
  \omega_d\big(D\varphi^t(u)\big) \leq
  \omega_d\big(\big(Dh(\varphi^t(u))\big)^{-1}\big)
  \omega_d\big(D\varphi_h^t(h(u))\big)
  \omega_d\big(Dh(u)\big)
\]
and
\[
  %\sup_{u \in K}
  \omega_d\big(D\varphi^t(u)\big) \leq
  c^2
  %\sup_{u \in K}
  \omega_d\big(D\varphi_h^t(h(u))\big).
\]
Therefore for any $d \in [0,n]$, $t\geq0$, and $u \in K$ we have
\begin{equation}\label{Omega2Horn}
  c^{-2}
  %\sup_{u \in K}
  \omega_d\big(D\varphi_h^t(h(u))\big)
  \leq
  %\sup_{u \in K}
  \omega_d\big(D\varphi^t(u)\big)
  \leq
  c^2
  %\sup_{u \in K}
  \omega_d\big(D\varphi_h^t(h(u))\big)
\end{equation}
and
\[
  (c^{-2}-1)
  %\sup_{u \in K}
  \omega_d\big(D\varphi_h^t(h(u))\big)
  \leq
  %\sup_{u \in K}
  \omega_d\big(D\varphi^t(u)\big)
  -
  %\sup_{u \in K}
  \omega_d\big(D\varphi_h^t(h(u))\big)
  \leq
  (c^2-1)
  %\sup_{u \in K}
  \omega_d\big(D\varphi_h^t(h(u))\big).
\]
If for $t\geq0$ there exists $d \in [0,n]$ such that
$\sup_{u \in K}\omega_d\big(D\varphi_h^t(h(u))\big) <1$ (see \eqref{wDphiht<1}),
then by \eqref{liminf} we get
\[
  \liminf\limits_{t \to +\infty}\omega_d\big(D\varphi_h^t(h(u))\big)=0
\]
and
\[
  0 \leq
  \liminf\limits_{t \to +\infty}
  \left(
  \omega_d\big(D\varphi^t(u)\big)
  -
  \omega_d\big(D\varphi_h^t(h(u))\big)
  \right)
  \leq
  0.
\]
$\blacksquare$
\begin{corollary}\label{thm:hdiffLE} (see, e.g. \cite{KuznetsovAL-2016})
For $u \in K$ we have
\[
  \lim\limits_{t \to +\infty}
  \bigg( \LEs_i\big( D\varphi_h^t(h(u)) \big) - \LEs_i\big(D\varphi^t(u)\big)\bigg)=0,
  \quad \quad i=1,2,..,n,
 \]
 and, therefore,
 \[
  \limsup\limits_{t \to +\infty} \LEs_i\big( D\varphi_h^t(h(u)) \big)
  =
  \limsup\limits_{t \to +\infty} \LEs_i\big( D\varphi^t(u) \big),
  \quad \quad i=1,2,..,n.
 \]
\end{corollary}
{\bf Proof.}
From \eqref{Omega2Horn} for $t>0$ we obtain
\begin{equation}\label{LE2Horn}
  \frac{1}{t}\ln c^{-2} +
  \frac{1}{t}\ln\omega_d\big(D\varphi_h^t(h(u))\big)
  \leq
  \frac{1}{t}\ln\omega_d\big(D\varphi^t(u)\big)
  \leq
  \frac{1}{t}\ln c^2 +
  \frac{1}{t}\ln\omega_d\big(D\varphi_h^t(h(u))\big).
\end{equation}
Thus for the integer $d=m$ we have
\[
  \lim\limits_{t \to +\infty}
  \left(
  \frac{1}{t}\ln\omega_m\big(D\varphi^t(u)\big)
  -
  \frac{1}{t}\ln\omega_m\big(D\varphi_h^t(h(u))\big)
  \right)
  =
  \lim\limits_{t \to +\infty}
  \left(
  \sum_{i=1}^{m} \LEs_i\big( D\varphi^t(u) \big)
  -
  \sum_{i=1}^{m} \LEs_i\big( D\varphi_h^t(h(u)) \big)
  \right)
  = 0.
\]
$\blacksquare$

The above statements are reformulations from \cite{KuznetsovAL-2016,LeonovAK-2015}
and imply the following

\begin{proposition}\label{thm:dDOunderdiff}
The Lyapunov dimension of the dynamical system $\{\varphi^t\}_{t\geq0}$
with respect to the compact invariant set $K$
is invariant with respect to any diffeomorphism
$h: U \subseteq \mathbb{R}^n \to \mathbb{R}^n$, i.e.
\begin{equation}\label{dDOunderdiff}
\begin{aligned}
  &
  \dim_{\rm L}(\{\varphi^t\}_{t\geq0},K)
  =
  \dim_{\rm L}(\{\varphi_h^t\}_{t\geq0},h(K)).
\end{aligned}
\end{equation}
\end{proposition}
{\bf Proof.}
Lemma~\ref{thm:hdiff} implies that
if
$
  \max_{w \in h(K)}\omega_d\big(D\varphi_h^{t}(w)\big) < 1
$
for $t>0$ and $d \in [0,n]$,
then there exists $T>t$ such that
\begin{equation}\label{defT}
  \max_{u \in K}\omega_d\big(D\varphi^{T}(u)\big) < 1
\end{equation}
and vice verse.
Thus, from \eqref{dimLest}, we have  
$\dim_{\rm L}(\{\varphi^t\}_{t\geq0},K) < d
\Leftrightarrow 
\dim_{\rm L}(\{\varphi_h^t\}_{t\geq0},h(K))<d$.
%Thus the set of $d$
%over which $\inf_{t>0}$ is taken in \eqref{DOmapt}
%is the same for $D\varphi^t(u)$ and $D\varphi_h^t(w)$ and, 
%Therefore,
%\[
%  \inf_{t>0}\inf\{d\in[0,n]: \max\limits_{u\in K}\omega_{d}(D\varphi^t(u))<1\}
%  =
%  \inf_{t>0}\inf\{d\in[0,n]: \max\limits_{w\in h(K)}\omega_{d}(D\varphi_h^t(w))<1\}.
%\]
$\blacksquare$

Remark that the invariance with respect to Lipschitz diffeomorphisms
is an essential point for the introduction of the Lyapunov dimension on manifolds.

\begin{corollary}\label{thm:hdiffLE}
Suppose $H(u)$ is an $\,n\times n $ matrix, all elements of which are scalar continuous functions of $u$, and $\det H(u) \neq 0$ for all $u \in K$.
If for $t>0$  there exists $d \in (0,n]$ such that
\begin{equation}\label{wDphiH<1}
  %\max_{w \in h(K)}\omega_d\big(D\varphi_h^t(w)\big)
  %=
  \max_{u \in K}
  \omega_d\bigg(
  H(\varphi^t(u))
  D\varphi^t(u)
  \big(H(u)\big)^{-1}
  \bigg) <1,
\end{equation}
then by \eqref{wDphiht<1} with $Dh(u)=H(u)$,
\eqref{dDOunderdiff}, and \eqref{defT}, for all sufficiently large $T>0$
we have
\[
   \dim_{\rm H}K \leq \dim_{\rm L}(\{\varphi^t\}_{t\geq0},K)
   \leq \dim_{\rm L}(\varphi^T,K) < d.
\]
If we take $H(u)=p(u)S$,
where $p: U \subseteq \mathbb{R}^n \to \mathbb{R}^1$
is a continuous positive scalar function, $p(u) \neq 0$ for all $u \in K \subset U$,
and $S$ is a nonsingular $n \times n$ matrix,
condition \eqref{wDphiH<1}
takes the form
\begin{equation}\label{wDphip<1}
  \sup_{u \in K}
  \omega_d\bigg(
  H(\varphi^t(u))
  D\varphi^t(u)
  \big(H(u)\big)^{-1}
  \bigg)
  =
  \sup_{u \in K}
  \bigg(
  \big(p(\varphi^t(u))p(u)^{-1}\big)^d
  \omega_d\big(SD\varphi^t(u)S^{-1}\big)
  \bigg)
  <1.
\end{equation}
\end{corollary}

Remark that if a nonsingular matrix $S$ is such that
\[
  SD\varphi^{t}(u)S^{-1}={\rm diag}(\lambda_1(t,u),..,\lambda_n(t,u)),
  \quad
  |\lambda_1(t,u)|\geq..\geq|\lambda_n(t,u)|,
\]
then $\sigma_i\big(SD\varphi^{t}(u)S^{-1}\big) =|\lambda_i(t,u)|$.

Let us apply the linear change of variables
$w=h(u)=Su$ with a nonsingular $n \times n$ matrix $S$.
Then
$\varphi^t(u_0) = u(t,u_0)$
is transformed into $\varphi_S^t(w_0)$
\[
   \varphi_S^t(w_0) = w(t,w_0)=S\varphi^t(u_0)=Su(t,S^{-1}w_0).
\]
Consider transformed
systems \eqref{eq:ode} and \eqref{eq:odife}
\[
  \dot w = Sf(S^{-1}w)
  \mbox{\ \ or \ }
  w(t+1) = S\varphi(S^{-1}w(t))
\]
and their linearizations along the solution
$\varphi_S^t(w_0)=w(t,w_0)=S\varphi^t(u_0)$:
\begin{equation}\label{jacobian-new}
\begin{aligned}
  &
  \dot v = J_S(w(t,w_0))v
  \mbox{\ \ \it or \ \ }v(t+1) = J_S(w(t,w_0))v(t),
 \\ &
  J_S(w(t,w_0)) =S \, J(S^{-1}w(t,w_0)) \, S^{-1}
  = S \, J(u(t,u_0)) \, S^{-1}.
\end{aligned}
\end{equation}
For the corresponding fundamental matrices we have $D\varphi_S^t(w)=SD\varphi^t(u)S^{-1}$.

\begin{proposition}\label{ueqcr}
Suppose that at one of the equilibrium points of the dynamical system $\{\varphi^t\}_{t\geq0}$:
$u_{eq}\equiv\varphi^t(u_{eq})$, $u_{eq} \in U$,
the matrix $J(u_{eq})$ has simple real eigenvalues:
$\{\lambda_i(u_{eq})\}_{i=1}^{n}$. Consider a nonsingular matrix $S$ such that
\[
  SJ(u_{eq})S^{-1}={\rm diag}\big(\lambda_1(u_{eq}),..,\lambda_n(u_{eq})\big),
\]
where
$\lambda_{i}(u_{eq}) \geq \lambda_{i+1}(u_{eq})$ for the case of continuous-time dynamical systems
and
$|\lambda_{i}(u_{eq})| \geq |\lambda_{i+1}(u_{eq})|$ for discrete-time dynamical systems.
Then
\[
  \dim_{\rm L}(\{\varphi^t\}_{t\geq0},u_{eq}) =
  \dim_{\rm L}(\{\varphi_S^t\}_{t\geq0},Su_{eq}) = \dim_{\rm L}(\varphi_S^t,Su_{eq}), \quad \forall t>0.
\]
Also, if for a certain $t=t^{cr}>0$
the maximum of finite-time local Lyapunov dimensions
$\dim_{\rm L}(\varphi_S^{t^{cr}},w)$
is achieved\,\,\footnote{
In general, since the function $u \mapsto \dim_{\rm L}(\varphi^t,u)$
is upper semi-continuous (see, e.g. \cite[p.554]{Gelfert-2003}),
there exists a critical point $u_{\rm L}(t) \in K$ (it may be not unique)
such that
$\sup_{u \in K}\dim_{\rm L}(\varphi^t,u)=\dim_{\rm L}(\varphi^t,u_{\rm L}(t))$.
An interesting question is
whether there exists a critical path
$\gamma^{cr}=\{\varphi^t(u_{\rm L}(t_0)),\ t\geq0 \}$
such that for each $t\geq0$
one of the corresponding critical points belongs to the critical path:
$\varphi^t(u_{\rm L}(t_0)) = u_{\rm L}(t)$,
and, if so, whether the critical path is
an equilibrium or a periodic solution.
The last part of the question was formulated
in \cite[p.98, Question 1]{Eden-1989-PhD}.
Taking into account \eqref{DOinctT}
we can consider increasing sequence $t_k \to +\infty$
such that $\dim_{\rm L}(\varphi^{t_k},u_{\rm L}(t_k))$
is monotonically decreasing to
$\inf_{t>0}\dim_{\rm L}(\varphi^t,u_{\rm L}(t))=\dim_{\rm L}(\{\varphi^t\}_{t\geq0},K)$.
Since $K$ is a compact set, we can obtain
a subsequence $t_m = t_{k_m} \to +\infty$ such that
there exists a limit critical point $u^{cr}_{\rm L}$:
$u_{\rm L}(t_m) \to u^{cr}_{\rm L} \in K$ as $m \to +\infty$.
Thus we have
$\dim_{\rm L}(\varphi^{t_m},u_{\rm L}(t_m)) \searrow
\dim_{\rm L}(\{\varphi^t\}_{t\geq0},K)$ and
$u_{\rm L}(t_m) \to u^{cr}_{\rm L} \in K$
as $m \to +\infty$.
}
at the point $w^{cr}_{eq}=Su_{eq} \in SK$:
\begin{equation}\label{DOeqpoint}
   \dim_{\rm L}(\varphi_S^{t^{cr}},w^{cr}_{eq}) =
   \sup_{w \in SK}\dim_{\rm L}(\varphi_S^{t^{cr}},w),
   %\quad
   %\varphi^t(u^{cr}_{eq})\equiv u^{cr}_{eq},
\end{equation}
then
\begin{equation}\label{DOeqpoint}
   \dim_{\rm H} K \leq
   \dim_{\rm L}(\{\varphi^t\}_{t\geq0},K) =
   \dim_{\rm L}(\{\varphi_S^t\}_{t\geq0},SK)
   = \dim_{\rm L}(\varphi_S^{t^{cr}},K)=\dim_{\rm L}(\varphi_S^{t^{cr}},w^{cr}_{eq}).
\end{equation}
Here  $\dim_{\rm L}(\varphi_S^{t^{cr}},w^{cr}_{eq})=\dim_{\rm L}(\varphi_S,w^{cr}_{eq})$
and it can be computed
by \eqref{lftKY} with
$\LEs_i(SJ(u_{eq})S^{-1})=\lambda_i(u_{eq})$ for continuous-time dynamical systems:
$\dim_{\rm L}(\varphi_S^{t^{cr}},w^{cr}_{eq})= \dim_{\rm L}^{\rm KY}(\{\lambda_i(u_{eq})\}_1^n)$,
and with $\LEs_i(SJ(u_{eq})S^{-1})=\ln|\lambda_i(u_{eq})|$
for discrete-time dynamical systems:
$\dim_{\rm L}(\varphi_S^{t^{cr}},w^{cr}_{eq})= \dim_{\rm L}^{\rm KY}(\{\ln|\lambda_i(u_{eq})|\}_1^n)$.
\end{proposition}
This statement may be useful in the numerical calculation of the Lyapunov dimension
for global attractors and B-attractors
(which contain unstable equilibria)
\cite{Chueshov-2015,LeonovKM-2015-EPJST}.

\bigskip
Consider now \emph{the Leonov method} of analytical estimation of the Lyapunov dimension
and its connection with the invariance of the Lyapunov dimension with respect to diffeomorphisms.
Following \cite{Leonov-1991-Vest,LeonovL-1993,Leonov-2002}, we consider a special class of diffeomorphisms
such that $Dh(u)=p(u)S$,
where $p: U \subseteq \mathbb{R}^n \to \mathbb{R}^1$
is a continuous scalar function and $S$ is a nonsingular $n\times n$ matrix.
Below it will be shown that the multiplier of the type $p(\varphi^t(u))(p(u))^{-1}$ in \eqref{wDphip<1}
plays the role of the Lyapunov-like functions.
This multiplier can also be interpreted as the changes of Riemannian metrics \cite{NoackR-1996}.

Consider continuous-time dynamical systems.
Let $\lambda_i(u_0,S)=\lambda_i(S\varphi^t(u_0))$, $i=1,2,...,n,$
be the eigenvalues of the symmetrized Jacobian matrix %\eqref{jacobian-new}
\begin{equation}\label{SJS}
 \frac{1}{2} \left( S J(u(t,u_0)) S^{-1} +
 (S J(u(t,u_0)) S^{-1})^{*}\right)
 =
 \frac{1}{2} \left(
 J_S(w(t,w_0))+J_S(w(t,w_0))^{*}
 \right),
\end{equation}
ordered so that
$\lambda_1(u_0,S)\geq \cdots \geq \lambda_n(u_0,S)$ for any $u_0 \in U$.
The following theorem is reformulation
of results from \cite{Leonov-2002,Leonov-2008,Leonov-2012-PMM}.

\begin{theorem}\label{thm:LD-estimate-V}
%Let $d=(j+s) \in [1,n]$, where
%integer $j=\lfloor d \rfloor \in \{1,\ldots,n\}$
%and real $s = (d - \lfloor d \rfloor) \in [0,1)$.
If there exist an integer $j \in \{1,\ldots,n-1\}$,
a real $s \in [0,1]$, a differentiable scalar function $V: U \subseteq \mathbb{R}^n \to \mathbb{R}^1$, and a nonsingular $n\times n$ matrix $S$
such that
\begin{equation}\label{ineq:weilSVct}
  \sup_{u \in K} \big(
  \lambda_1 (u,S) + \cdots + \lambda_j (u,S)
  + s\lambda_{j+1}(u,S) + \dot{V}(u)
  \big) < 0,
\end{equation}
where $\dot{V} (u) = ({\rm grad}(V))^{*}f(u)$,
then
\[
   \dim_{\rm H}K \leq
    \dim_{\rm L}(\{\varphi^t\}_{t\geq0},K)
    \leq
    \dim_{\rm L}(\varphi^T,K)
   < j+s
\]
for all sufficiently large $T>0$.
\alarm{} %(\cite[p.262]{BoichenkoLR-2005}: $j \in \{1,\ldots,n-1\}$ $\dim_{\rm H} K  \leqslant j+s$)
\end{theorem}
{\bf Proof}.
Let $p(u) = e^{V(u)(j+s)^{-1}}$.
From the following relation \cite{Smith-1986}
(see, also \cite{BoichenkoLR-2005}[p.48]) %,\cite[p.460]{LeonovPS-1996}
\begin{equation}\label{weil}
\begin{aligned}
  &
  \omega_{j+s}\big(SD\varphi^t(u)S^{-1}\big)
  \leq
   \exp\left(
     \int_0^t
     \lambda_1(S\varphi^\tau(u))+\cdots+\lambda_j(S\varphi^\tau(u))+s\lambda_{j+1}(S\varphi^\tau(u))
      d\tau
   \right)
\end{aligned}
\end{equation}
and the relation
\[
  \big(p(\varphi^t(u))p(u)^{-1}\big)^{j+s}
  =
  \exp\big( V(\varphi^t(u))-V(u)\big)=
  \exp\left( \int_{0}^{t} \dot V(\varphi^\tau(u)) d\tau \right)
\]
we get
\begin{equation}
\begin{aligned}
  &
  \big(p(\varphi^t(u))p(u)^{-1}\big)^{j+s}
  \omega_{j+s}\big(SD\varphi^t(u)S^{-1}\big)
  \leq
  \\
  & \qquad
  \leq
   \exp\left(
     \int_0^t \big(
     \lambda_1(S\varphi^\tau(u))+\cdots+\lambda_j(S\varphi^\tau(u))+s\lambda_{j+1}(S\varphi^\tau(u))
     +\dot V(\varphi^\tau(u))
     \big) d\tau
   \right).
\end{aligned}
\end{equation}
Since $\varphi^t(u) \in K$ for any $u \in K$,
by \eqref{ineq:weilSVct} for $t>0$ we have
\[
  \max_{u \in K} \bigg(
  \big(p(\varphi^t(u))p(u)^{-1}\big)^{j+s}
  \omega_{j+s}\big(SD\varphi^t(u)S^{-1}\big)\bigg)
  < 1, \quad t>0.
\]
Therefore by Corollary~\ref{thm:hdiffLE} with
$H(u) = p(u)S$,
where
$p(u) = \left(e^{V(u)}\right)^{\frac{1}{j+s}}$,
we get the assertion of the theorem.
$\blacksquare$

{\bf Remark}.
The idea of the estimation of the Hausdorff dimension
by the eigenvalues of symmetrized Jacobian matrix
was developed in
\cite{DouadyO-1980,Smith-1986}
(e.g. for $\dot V(u) \equiv 0$, condition \eqref{ineq:weilSVct}
is considered in \cite{Smith-1986}).
The function $p(u)=e^{V(u)(j+s)^{-1}}$ was introduced in \cite{Leonov-1991-Vest}
($\dot{V}(u)$ allows one to effectively estimate the partial sum of the eigenvalues)
and matrix $S$ was introduced in \cite[eq.(8)]{LeonovL-1993}
(for the simplification of eigenvalues computation).
Condition \eqref{ineq:weilSVct} is valid
if $\dot V(u)$ is continuous and
$
\lambda_1 (u,S) + \cdots + \lambda_j (u,S) + s\lambda_{j+1}(u,S) + \dot{V}(u) < 0
$ for all $u \in K$ (here we take into account that $J(u,S)$ and, thus, $\lambda_i(u,S)$ are continuous).
The constancy of the signs of $V(u)$ or $\dot V(u)$ is not required in the theorem.

\medskip
Next, we consider discrete-time dynamical systems.
Let $\lambda_i(u_0,S)=\lambda_i(S\varphi^t(u_0))$, $i=1,2,...,n$
be the positive square roots of the eigenvalues of the symmetrized Jacobian matrix
(i.e. the singular values of the Jacobian matrix)
\begin{equation}\label{SJS}
 \left( (S J(u(t,u_0)) S^{-1})^{*}S J(u(t,u_0)) S^{-1} \right)
 =
 \left(
 J_S(w(t,w_0))^{*}J_S(w(t,w_0))
 \right),
\end{equation}
ordered so that
$\lambda_1(u_0,S)\geq \cdots \geq \lambda_n(u_0,S)>0$ for any $u_0 \in U$.

\begin{theorem}\label{thm:LD-estimate-Vdt}
%Let $d=(j+s) \in [1,n]$, where
%integer $j=\lfloor d \rfloor \in \{1,\ldots,n\}$
%and real $s = (d - \lfloor d \rfloor) \in [0,1)$.
If there exist an integer $j \in \{1,\ldots,n-1\}$, 
a real $s \in [0,1]$,
a continuous scalar function $V: K \subseteq \mathbb{R}^n \to \mathbb{R}^1$,
and a nonsingular $n\times n$ matrix $S$
such that
\begin{equation}\label{ineq:weilSV}
  \sup_{u \in K}\bigg(
  \ln\lambda_1 (u,S) + \cdots + \ln\lambda_j (u,S)
  + s\ln\lambda_{j+1}(u,S) + \big(V(\varphi(u))-V(u)\big)
  \bigg) < 0,
\end{equation}
then
\[
   \dim_{\rm H}K \leq
    \dim_{\rm L}(\{\varphi^t\}_{t\geq0},K)
    \leq
    \dim_{\rm L}(\varphi^T,K)
   < j+s
\]
for all sufficiently large $T>0$.
\end{theorem}
{\bf Proof}.
By \eqref{horn}
for
$D\varphi_S^t(w)=SD\varphi^t(u)S^{-1} =
\prod\limits_{\tau=0}^{t-1}\big(S \, J(u(\tau,u_0)) \, S^{-1}\big)$
we have
\begin{equation}\label{weilds}
\begin{aligned}
  &
  \omega_{j+s}\big(SD\varphi^t(u)S^{-1}\big)
  \leq
  \prod_{\tau=0}^{t-1} \omega_{j+s}\big(S \, J(u(\tau,u_0)) \, S^{-1}\big).
  %=
  %\prod_{\tau=0}^{t-1} \lambda_1(S\varphi^\tau(u))\cdots\lambda_j(S\varphi^\tau(u))\lambda_{j+1}(S\varphi^\tau(u))^s.
\end{aligned}
\end{equation}
Therefore we get (the discrete analog of \eqref{weil})
\begin{equation}
  \omega_{j+s}\big(SD\varphi^t(u)S^{-1}\big)
  \leq
  \prod_{\tau=0}^{t-1} \lambda_1(S\varphi^\tau(u))\cdots\lambda_j(S\varphi^\tau(u))\big(\lambda_{j+1}(S\varphi^\tau(u))\big)^s.
\end{equation}
Let $p(u) = e^{V(u)(j+s)^{-1}}$.
By the relation
\[
  \big(p(\varphi^t(u))p(u)^{-1}\big)^{j+s}
  =
  \exp\big( V(\varphi^t(u))-V(u)\big)
  =
  \exp\bigg( \sum_{\tau=0}^{t-1} V(\varphi^{\tau+1}(u))-V(\varphi^{\tau}(u)) \bigg)
\]
we obtain
\[
\begin{aligned}
  &
  \ln\big(p(\varphi^t(u))p(u)^{-1}\big)^{j+s}
  +\ln\omega_{j+s}\big(SD\varphi^t(u)S^{-1}\big)
  \leq
  \\
  & \qquad
  \leq\sum_{\tau=0}^{t-1}\bigg(
  \ln\lambda_1(S\varphi^\tau(u))+\cdots+\ln\lambda_j(S\varphi^\tau(u))+s\ln\lambda_{j+1}(S\varphi^\tau(u))
  +V(\varphi(\varphi^{\tau}(u)))-V(\varphi^{\tau}(u))\bigg).
\end{aligned}
\]
Since $\varphi^t(u) \in K$ for any $u \in K$, by \eqref{ineq:weilSV}
and Corollary~\ref{thm:hdiffLE} with $H(u) = p(u)S$, where
$p(u) = \left(e^{V(u)}\right)^{\frac{1}{j+s}}$,
we get the assertion of the theorem.
$\blacksquare$

\medskip

\begin{corollary}
  If conditions of Theorems~\ref{thm:LD-estimate-V} or \ref{thm:LD-estimate-Vdt}
  are valid for all $\overline{d}=(j+s) \in (d,n]$, 
  and at an equilibrium point $u^{cr}_{eq} \equiv \varphi^{t}(u^{cr}_{eq})$
  the relation
  \[
    \dim_{\rm L}(\{\varphi^t\}_{t\geq0},u^{cr}_{eq}) = d
  \]
  holds, then
  for any invariant set $K \supset u^{cr}_{eq}$
  from \eqref{DOeqpoint}
  we get the formula of exact Lyapunov dimension
  \[
  \begin{aligned}
   & \dim_{\rm H}K \leq
   \dim_{\rm L}(\{\varphi^t\}_{t\geq0},K)
   =\dim_{\rm L}(\{\varphi^t\}_{t\geq0},u^{cr}_{eq})=d.
  \end{aligned}
  \]
\end{corollary}
In \cite{BoichenkoL-1998,PogromskyM-2011} it is demonstrated how a technique similar to the above can be effectively used to derive constructive upper bounds of the topological entropy of dynamical systems.

\bigskip
{\bf Example.}
Consider the Henon map $\varphi_{\rm Henon} \colon \mathbb{R}^2\to\mathbb{R}^2$
\begin{equation}\label{henonmap}
 \left(
 \begin{array}{l}
   x\\ y
 \end{array}
 \right)
 \mapsto
 \left(
 \begin{array}{l}
   a+by-x^2 \\ x
 \end{array}
 \right),
\end{equation}
where $a>0$, $b\in(0,1)$ are the parameters of mapping.
The stationary points $(x_\pm,x_\pm)$ of this map are the following:
$x_\pm=\frac{1}{2} \,\big(b-1\pm\sqrt{(b-1)^2+4a}\big)$.
Following \cite{Leonov-2002} we consider
\[
S =
\left(
  \begin{array}{cc}
    1 & 0 \\
    0 & \sqrt{b} \\
  \end{array}
\right),
\gamma= \frac{1}{(b-1-2x_{-})\sqrt{x_{-}^2+b}}, \ s \in [0,1).
\]
In this case
\[
SJ\big((x,y)\big)S^{-1} =
\left(
  \begin{array}{cc}
    -2x & \sqrt{b} \\
    \sqrt{b} & 0 \\
  \end{array}
\right), \
\lambda_1((x,y),S) = \big(\sqrt{x^2+b}+|x|\big), \
\lambda_2((x,y),S) = \frac{b}{\lambda_1((x,y),S)}.
\]
If we take $V((x,y)) = \gamma(1-s)(x+by)$,
then condition \eqref{ineq:weilSV} with $j=1$
and
\[
s>s^*=\dfrac{\ln|\lambda_1((x_{-},x_{-}),S)|}{|\ln b- \ln|\lambda_1((x_{-},x_{-}),S)||}
\]
is satisfied for all $(x,y) \in \mathbb{R}^2$
and we need not localize the set $K$ in the phase space.
By Proposition~\ref{ueqcr} and \eqref{lftKY}, at the equilibrium point $u^{cr}_{eq} = (x_{-},x_{-})$
we have
\[
  \dim_{\rm L}(\{\varphi^t_{\rm Henon}\}_{t\geq0},(x_{-},x_{-})) =
  %\dim_{\rm L}(S\varphi^{t^{cr}}_{\rm Henon},w^{cr}_{eq})=
  \dim_{\rm L}^{\rm KY}(\{\ln\lambda_i(x_{-},x_{-})\}_1^2)
  =
  1+s^*.
\]
Therefore, for a bounded invariant set $K^{B}_{\rm Henon} \ni (x_-,x_-)$
(e.g. B-attractor)
we have \cite{Leonov-2002}
$$
  \dim_{\rm L}(\{\varphi^t_{\rm Henon}\}_{t\geq0},K^{B}_{\rm Henon}) =
  \dim_{\rm L}(\{\varphi^t_{\rm Henon}\}_{t\geq0},(x_{-},x_{-}))
  =1+\dfrac{\ln|\lambda_1((x_{-},x_{-}),S)|}{|\ln b- \ln|\lambda_1((x_{-},x_{-}),S)||}.
$$
Here for $a=1.4$ and $b=0.3$ we have
$\dim_{\rm L}(\{\varphi^t_{\rm Henon}\}_{t\geq0},K^{B}_{\rm Henon})=1.495\,...$.
Remark that numerical localization of attractors\footnote{
 The classical Henon attractor is not global, but it is \emph{self-excited}
 with respect to both equilibria %$(x_{-},x_{-})$
 and, thus, can be visualized numerically by a trajectory with the initial data in their vicinities
 (i.e. the basin of attraction intersects with small neighborhoods of the equilibria;
 if there are no such equilibria, the attractor is called a \emph{hidden attractor}
 \cite{LeonovKV-2011-PLA,LeonovK-2013-IJBC,LeonovKM-2015-EPJST,Kuznetsov-2016}).
 Hidden attractors can be found in the Henon map with negative $b$ 
 (see, e.g. $a= 1.63$, $b=-0.138$).
 A \emph{conjecture on the Lyapunov dimension of self-excited attractors} 
 is that \emph{for ''typical'' systems the Lyapunov dimension of self-excited attractor
 is less then the Lyapunov dimension of one of the unstable equilibria,
 the unstable manifold of which intersects with the basin of attraction
 and allows to visualize the attractor}. 
} %$K_{\rm Henon}$
%$K_{\rm Henon}$ and $K^{B}_{\rm Henon}$
by the square: $-1.8 \leq x,y \leq 1.8$,
allows one to estimate directly the singular values of $J(x,y)$
and obtain the estimation $\dim_{\rm L}(\{\varphi^t_{\rm Henon}\}_{t\geq0},K^{B}_{\rm Henon}) \leq 1.523$ \cite{Hunt-1996}
(this approach corresponds to the use of $V(u)\equiv1$ in \eqref{ineq:weilSV}).
For attractor $K_{\rm Henon}$ (which does not include equilibria)
it is known from numerical experiments \cite{RusselHO-1980}
that
$\dim_{\rm H}K_{\rm Henon} \approx 1.261$
is in good agreement\footnote{
%It is not difficult to construct an artificial example
%where these dimensions do not coincide.
%E.g. one may consider the tent map
%$x_{k+1}=\left\{
%\begin{array}{ll}
%  3x & x < 1/2 \\
%  3(1-x) & x \geq 1/2
%\end{array}
%\right.
%$, $x \in [0,1]$,
%which generates canonical Cantor set with the Hausdorff dimension $\ln2/\ln3$,
%and add the equation $y_{k+1} = a^{-1} y_{k}$, $|a| \geq 3$, $y \in [0,1]$.
%Here, the Lyapunov dimension of considered two-dimensional map is $(1+\ln3/\ln a)$.
%See also examples in \cite{Sell-1989}.
Examples are known where such a good agreement
between these dimensions does not take place \cite{Sell-1989}.
}
with $\dim_{\rm L}(\{\varphi^t_{\rm Henon}\}_{t\geq0},K_{\rm Henon}) \approx 1.264$;
algorithm from \cite{LeonovKM-2015-EPJST}
gives $\dim_{\rm L}(\{\varphi^t_{\rm Henon}\}_{t\geq0},K_{\rm Henon}) \approx 1.262$.

\section{Conclusion}

In this work the Leonov method, based on the direct Lyapunov method
with special Lyapunov-like functions, is derived from
the invariance of the Lyapunov dimension with respect to diffeomorphisms.
The advantage of this method is that it allows one to estimate the Lyapunov dimension
of invariant sets without localization of the set in the phase space
and, in some cases, to get effectively
exact Lyapunov dimension formulas
\cite{Leonov-2002,LeonovP-2005,LeonovPS-2011,LeonovK-2015-AMC,LeonovKKK-2015-arXiv-LuChen,LeonovKKK-2015-arXiv-YangTigan,LeonovAK-2015,LeonovKKK-2015-arXiv-Lorenz}
(another approaches for rigorous derivation of
exact Lyapunov dimension formulas are demonstrated, e.g. in \cite{DoeringGHN-1987,Eden-1990,Schmeling-1998}).

Remark that while the notion of local Lyapunov dimension is natural for the maps,
a rigorous introduction of this notion for the dynamical systems is a challenging task.
The above consideration is based
on the Douady-Oesterl\'{e} theorem
on the Lyapunov dimension of maps only.
The results on the Lyapunov dimension of dynamical systems,
developed by Constantin, Eden, Foia\c{s}, and Temam in \cite{ConstantinFT-1985,Eden-1990,EdenFT-1991},
have not been applied.
In the definition of the Lyapunov dimension of the dynamical system $\{\varphi^t\}_{t\geq0}$
(see \eqref{DOmaptmax} and \eqref{DOlim})
they consider
$\big(\omega_{d}(D\varphi^t(u))\big)^{1/t}$
instead of $\omega_{d}(D\varphi^t(u))$
and apply the theory of positive operators \cite{ChoquetF-1975}
(see also \cite{GundlachS-2000})
to prove the existence of a critical point\footnote{
See also the corresponding \emph{Eden conjecture} \cite[p.98, Question 1]{Eden-1989-PhD}.
From Fekete's lemma for the subadditive functions \cite[pp.463-464]{Kuczma-2009}
it follows that for any $d \in [0,n]$ there exists
$\lim_{t \to +\infty}\sup\limits_{u \in K}\big(\omega_{d}(D\varphi^t(u))\big)^{1/t}$.
Remark, that for the definition \eqref{DOmaptmax} with $\omega_{d}(D\varphi^t(u))$
the change of order of $\inf_{t > 0}$ and $\sup\limits_{u \in K}$ in \eqref{DOlim},
e.g. justification of
the use of $\sup\limits_{u \in K} \limsup\limits_{t \to +\infty} \dim_{\rm L}(\varphi^t,u)$
instead of $\liminf\limits_{t \to +\infty} \sup\limits_{u \in K} \dim_{\rm L}(\varphi^t,u)$,
is an open problem.
One might check the coincidence of $u^{cr}_{\rm L}$ and $u^{cr}_{\rm E}$.
} $u^{cr}_{\rm E}$ (which may be not unique),
where the corresponding global Lyapunov dimension
achieves maximum
(see \cite{Eden-1990}):
%\begin{equation}\label{gLDl}
\[
\begin{aligned}
  &
  \dim_{\rm L}^{\rm E}(\{\varphi^t\}_{t\geq0},K)
  &=&
  \inf\{d \in [0,n]:
  \lim\limits_{t \to +\infty} \max\limits_{u \in K}\ln\big(\omega_{d}(D\varphi^t(u))\big)^{1/t}<0\} =
  \\
  &&=&
  \inf\{d \in [0,n]:
  \limsup\limits_{t \to +\infty}\ln\big(\omega_{d}(D\varphi^t(u^{cr}_{\rm E}))\big)^{1/t}<0\}
  =
  \dim_{\rm L}^{\rm E}(\{\varphi^t\}_{t\geq0},u^{cr}_{\rm E}),
\end{aligned}
\]
and, thus, rigorously justify the use of the local Lyapunov dimension
$\dim_{\rm L}^{\rm E}(\{\varphi^t\}_{t\geq0},u)$.
%and $\sup_{u \in K}\dim_{\rm L}^{\rm E}(\{\varphi^t\}_{t\geq0},u)$.
%\end{equation}
%}

\section*{Acknowledgments}

The author would like to thank
Alp~Eden, Gennady~Leonov, Joseph~Oesterl\'{e}, Volker~Reitmann,
and James~Yorke for their comments.
The work was supported by Russian Science Foundation (project 14-21-00041)
and Saint-Petersburg State University.

%\section*{References}
%\bibliographystyle{apalike}
%\bibliography{C:/NICK/Work/!LEONOV/BIB/bib_nk,C:/NICK/Work/!LEONOV/BIB/bib_leonov,C:/NICK/Work/!LEONOV/BIB/bib_full,C:/NICK/Work/!LEONOV/BIB/bib_pll,C:/NICK/Work/!LEONOV/BIB/bib-gly}
%\bibliography{C:/Dropbox/bib/bib_nk,C:/Dropbox/bib/bib_leonov,C:/Dropbox/bib/bib_full,C:/Dropbox/bib/bib_pll,C:/Dropbox/bib/bib-2008-str-at}

\end{document}